\documentclass[12pt]{amsart}
\usepackage{color}
\usepackage{epsfig}
\usepackage{verbatim}
\usepackage{url}
\usepackage{parskip}

\usepackage{algorithm}
\usepackage{algpseudocode}
\usepackage{amssymb}

\newcommand{\algorithmicbreak}{\textbf{break}}
\newcommand{\Break}{\State \algorithmicbreak}

\let\bbordermatrix\bordermatrix
\patchcmd{\bbordermatrix}{8.75}{4.75}{}{}
\patchcmd{\bbordermatrix}{\left(}{\left[}{}{}
\patchcmd{\bbordermatrix}{\right)}{\right]}{}{}

\newtheorem{theorem}{Theorem}[section]
\newtheorem{lemma}[theorem]{Lemma}

\begin{document}

\title {On the epidemic threshold of a network}
\maketitle

\begin{center}
V. Cherniavskyi,\footnote{Department of Mathematics,
Brooklyn College, 2900 Bedford Avenue, Brooklyn, NY 11210.} 
G. Dennis,\footnote{Department of Mathematics,
Brooklyn College, 2900 Bedford Avenue, Brooklyn, NY 11210.}
S. R. Kingan \footnote{Department of Mathematics,
Brooklyn College, 2900 Bedford Avenue, Brooklyn, NY 11210,
and CUNY Graduate Center, 365 Fifth Avenue, New
York, NY 10016.}  
\end{center}

{\begin{abstract}

The graph invariant examined in this paper is the largest eigenvalue of the adjacency matrix of a graph.  Previous work demonstrates the tight relationship between this invariant, the birth  and death rate of a contagion spreading on the graph, and the trajectory of the contagion over time. We begin by conducting a simulation confirming this and explore bounds on the birth and death rate in terms of well-known graph invariants. As a result, the change in the largest eigenvalue resulting from removal of a vertex in the network is the best measure of effectiveness of interventions that slow the spread of a contagion.  We define the spread centrality of a vertex $v$ in a graph $G$ as the difference between the largest eigenvalues of $G$ and $G-v$. While the spread centrality is a distinct centrality measure and serves as another graph invariant for distinguishing graphs, we found  experimental evidence that vertices ranked by the spread centrality and those ranked by eigenvector centrality are strongly correlated. Since eigenvector centrality is easier to compute than the spread centrality, this  justifies using eigenvector centrality as a measure of spread, especially in large networks with unknown portions. We also examine two strategies for selecting members of a population to vaccinate.
\end{abstract}

\bigskip
 
\section {\bf  Introduction}

The well known Vertex Reconstruction Conjecture asks whether or not a graph $G$ with $n$ vertices $v_1, v_2, \dots , v_n$ can be reconstructed from its deck of vertex-deleted subgraphs $G-v_1, G-v_2, … , G-v_n$. This suggests a vertex centrality measure. The impact of vertex $v$ can be measured by removing it and considering the subgraph $G-v$. Then various graph invariants can be calculated for $G$ and compared with the corresponding parameters for $G-v$, thereby giving a  ranking of the vertices. Not every invariant has a practical application, but when it does, this approach gives a meaningful centrality measure. 

The parameter examined in this paper is the largest eigenvalue of the adjacency matrix of the graph. 
The inverse of the largest eigenvalue is the epidemic threshold in a non-linear dynamical system model of an epidemic spreading on a network of people or animals and removing a vertex corresponds to vaccinating the vertex. See for example \cite{Chak2008}, \cite{DaleyGani1999},  \cite{MooreRogers2020}, \cite{SarkarJalan2018}, and \cite{Rest2006}. 

The largest eigenvalue, also called the {\it index} or the {\it spectral radius}, is important  for several dynamic processes in addition to pandemics. For example, it is a key parameter in a virus spreading on a computer network, an idea spreading on a social network, channel capacity in Shannon information theory, energy levels of electrons in molecules, synchronization of coupled oscillators, stability of couplings in the  brain, etc. Hundreds of papers and multiple books have been written about the eigenvalues of a graph and their applications. See for example  \cite{Cvet2009} and \cite{Pik2003}.

In Section 2, we describe the epidemic threshold in detail and present the results of our simulation of a disease spreading on a network with various parameters designed to produce an outbreak that either grows to become an epidemic or dissipates. What does an epidemic that lingers in a population look like as vertices get infected, recover, and get infected again? What does an outbreak that dies out without becoming an epidemic look like? We present a way of visualizing these very different outcomes in the same figure. The theoretical results in this section establish relationships between graph parameters and epidemic parameters.

In Section 3 we define the spread centrality of a vertex in a graph $G$ as the difference between the largest eigenvalues of $G$ and $G-v$ and compare it to other centrality measures. Although this idea has apeared informally in \cite{Chak2008}, \cite{SarkarJalan2018}, and \cite{Rest2006}, this is the first time it has been formally defined as a centrality measure. It is the canonical centrality measure quantifying spread in the network. We compared it to other centrality measures with a surprising outcome. It is distinct from all known centrality measures, and thereby of theoretical importance in distinguising graphs. However, it does correlate with eigenvector centrality, thereby adding a previously unknown supporting argument to the premise in  \cite{Can2006}  that eigenvector centrality is a good measure of spread.  Finally, we also present two vaccination strategies and compare them for effectiveness. Through out the paper we focus on presenting rigourous results whereever possible and identified avenues for further mathematical research.


\section {\bf Epidemic Threshold}

There are two fundamental ways of modeling the spread of a disease, the SIS model and the SIR model. In the SIS model,  vertices have two states, susceptible (S) or infected (I). A susceptible  vertex becomes infected through an adjacent infected  vertex with probability $p_b$ called the birth rate of the virus. An infected  vertex recovers with probability $p_d$ called the death rate of the virus. In the SIS model a recovered  vertex can become infected again. It is assumed that a  vertex is infectious as soon as it becomes infected and becomes susceptible as soon as it recovers. In the SIR model,  vertices  have three states:  susceptible (S), infected (I), and removed (R). A susceptible  vertex becomes infected through an adjacent infected  vertex and recovers (or dies) and is removed from the network. For example, a disease like the flu follows the SIS model, whereas a disease like mumps where the patient becomes permanently immune after recovery follows the SIR model. 

These models when combined with information about the underlying network simulate the spread of diseases. Let $G$ be a connected graph with $n\ge 2$ vertices and $m$ edges. Suppose a virus is spreading on $G$, where $p_b$ is the probability that the virus spreads from one  vertex to a  vertex adjacent to it (the virus is born), and $p_d$ is the probability that an infected  vertex recovers (the virus dies). From epidemiological studies, these numbers are very small, usually less than $0.1$. 
An {\it epidemic threshold} is a number $\tau(G)$   such that if  $\frac{p_b}{p_d} > \tau(G),$ then the virus causes an epidemic, otherwise it dies out without causing an epidemic \cite {DaleyGani1999}. 

Let $A(G)$ be the adjacency matrix of graph $G$ and let the $n$ eigenvalues of $A(G)$ in descending order be 
$\lambda_1 (G)\ge \lambda_2(G) \ge  \cdots \ge \lambda_{n-1}(G) \ge \lambda_n(G).$  Since $A(G)$ is  a real symmetric matrix, the Perron-Frobenius Theorem implies that the largest eigenvalue $\lambda_1(G)$ is a real number and it has the highest absolute value of all the eigenvalues. If $G$ has no edges, then the adjacency matrix is the zero matrix, and the eigenvalues are 0. If $G$ has at least one edge, then $\lambda_1(G)>0$. If $G$ is connected, then $\lambda_1(G)>1$ is an eigenvalue of multiplicity 1 and all the entries of its eigenvector have the same sign. So we may consider the eigenvector to have all positive entries.

In \cite{Chak2008}, the authors developed a non-linear dynamical system (NLDS) model for virus propagation on a network based on the SIS model. A system is considered unstable if the first eigenvalue of a certain matrix $M$ associated with the system is large and stable otherwise. A small perturbation in a stable system will eventually die out.  
The authors showed that for a graph $G$ and a contagion with birth and death probabilities $p_b$ and $p_d$, respectively, spreading on the graph, the contagion will inevitably die out if and only if $\frac{p_b}{p_d} \le \frac{1}{\lambda_1(G)}$. Let ${\bf p}_t$ be a vector of length $n$ where the $i^{th}$ entry is the probability that vertex $i$ is infected at time $t$. If ${\bf p}_t = {\bf 0}$ for some $t$, then ${\bf p}_{t + k} = {\bf 0}$ for all $k \ge 0$. They showed that the sequence ${\bf p}_0, {\bf p}_1, \dots$ converges to $\bf 0$ if and only if $\frac{p_b}{p_d} \le \frac{1}{\lambda_1(G)}$, using a non-linear dynamical system relating $p_b$, $p_d$, $G$ and ${\bf p}_t$. Otherwise, if $\frac{p_b}{p_d} > \frac{1}{\lambda_1(G)}$ the epidemic does not die out.

\begin{theorem}\label{NLDS-theorem} Let $G$ be a graph and let $\lambda_1(G)$ be the largest eigenvalue of its adjacency matrix. In NLDS, the epidemic threshold $\tau(G) = \frac{1}{\lambda_1(G)}$. $\qed$
\end{theorem}

Consequently, the epidemic will die out over time irrespective of the size of the initial outbreak of infection, if $p_d > \lambda_1(G)p_b.$   For example, the graph on the left in Figure \ref{karategrapheigenvectorcentrality}, with 50 vertices and 250 edges, has largest eigenvalue 10.73 and therefore, if $p_d > 10.73p_b$, an epidemic spreading on the graph will die out over time. The graph on the right, with 50 vertices and 1185 edges, has largest eigenvalue 47.42, and requires $p_d > 47.42p_b$ for an epidemic to die out.

\begin{figure}[h]
\centering
\includegraphics[width=4in]{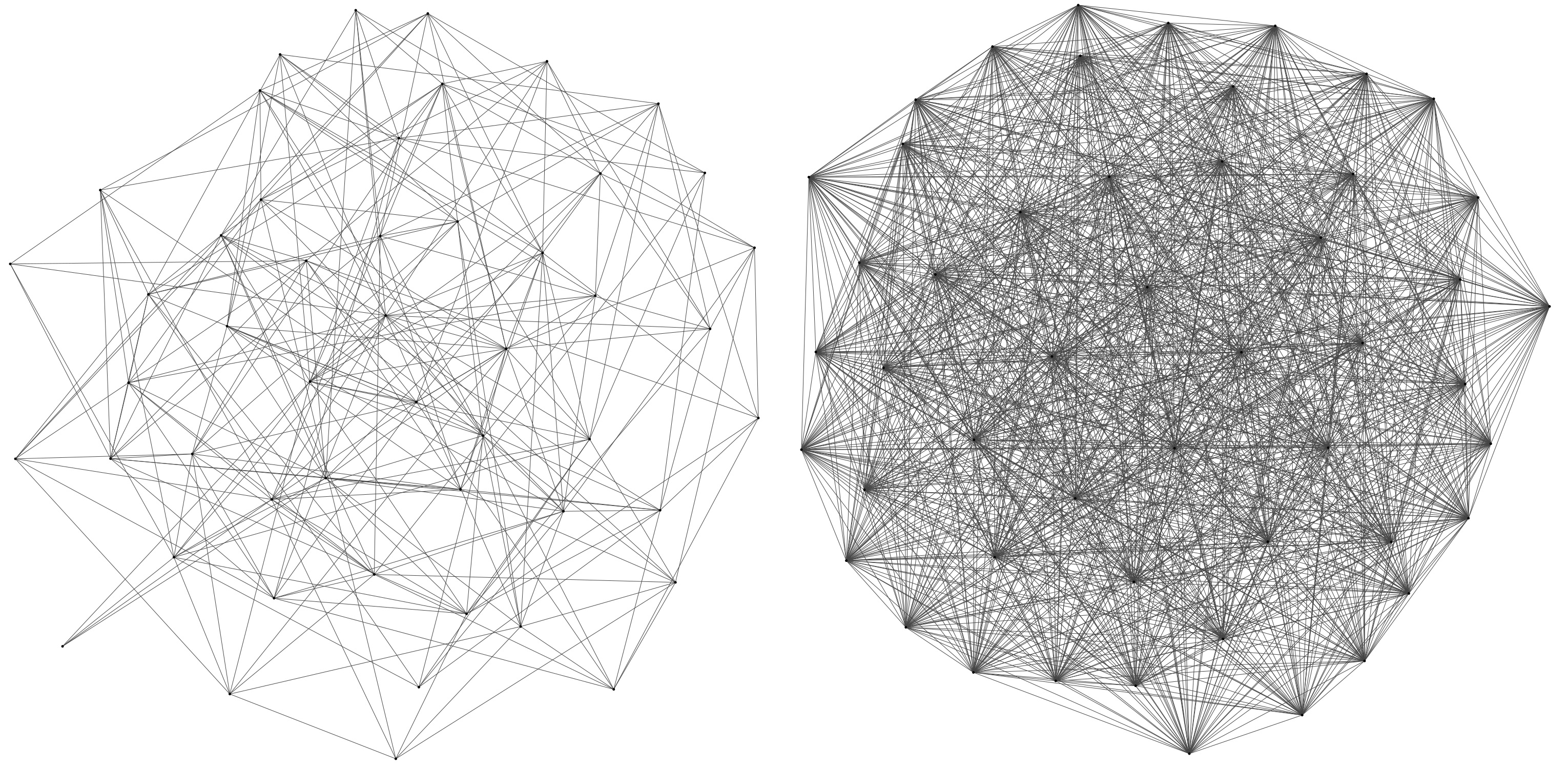}
\label{karategrapheigenvectorcentrality}
\caption {Two connected graphs on 50 vertices}
 \end{figure}


\section{\bf  Simulating an Epidemic}

Three items are key to the simulation: the birth rate $p_b$ of the virus, the death rate $p_d$ of the virus, and the underlying network $G$. The network gives $\lambda_1(G)$. The birth rate of the virus, $p_b$, is the probability that a vertex with one infected neighbor will become infected over a unit of time. If  vertex $v$ has $i$ infected neighbors, then perform $i$ random draws with probability $p_b$.  In other words, draw a random number uniformly between 0 and 1 and treat success as the event that the random number is less than $p_b$. Repeat this $i$ times because  vertex $v$ has $i$ infected neighbors. Each random draw is an independent event. If a positive result occurs in any draw, then $v$ is infected.

The probability of an infected  vertex recovering at each unit of time is the death rate of the virus $p_d$. The same random draw process is used to decide whether an infected  vertex will recover at each unit in time. In order to illustrate the relationship between $p_b$ and $p_d$, it makes sense to fix one of them and vary the other. Therefore, let $p_b$ be fixed at $0.05$ throughout the simulation.

Let $G$ be the well-known karate graph\footnote{\url{https://networkrepository.com/soc-karate.php}} described in \cite{Zachary1977}, which has 34 vertices and 78 edges, with $\lambda_1(G) = 6.73$. Suppose $p_d=0.4$, which is slightly larger than $p_b\lambda_1(G)$. The theory predicts that such an outbreak would die out on this graph.  One vertex $v$  is selected as the seed (patient zero) and fixed. The simulation is run for $T\ge 1$ days. For each day $t$,  where $2\le t\le T$:

\begin{itemize}
\item For each uninfected  vertex decide whether or not it gets infected at day $t$ based on the number of infected neighbors it has on day $t-1$. As mentioned above, this is done with a random draw with probability of success $p_b$ repeated as many times as the number of infected neighbors. 

\item For each infected  vertex decide whether or not it will recover with a single random draw with probability of success $p_d$. 

\end{itemize}
\noindent Finally, count the number of infected  vertices for each day. 

Next, in order to mitigate the effect of individual random number draws, this process is repeated $K$ times for seed  vertex $v$ with different sequences of random numbers governing the spread of the infection. We used $K=200$. 

At the end, for the seed  vertex $v$, let $S_{v, k, t}$ denote the number of infected  vertices, where $1\le t\le T$ is the number of days and $1\le k\le K$ is the index of the repetition. The entire process is repeated using each of the $n$  vertices in the graph as the seed  vertex in turn and the number of infected  vertices for each day $t$ across all seed vertices and all runs are averaged to get:
$$S_t = \frac{1}{nK} \sum_{v \in V(G)} \sum_{1 \le k \le K} S_{v, k, t}.$$
This gives a curve with days $t$ on the $x$-axis and the number of infected  vertices $S_t$ on the $y$-axis corresponding to $p_b=0.5$ and $p_d=0.4$ over all $T$ days. Individual curves $S_{t,v} = \frac{1}{K} \sum_{1 \le k \le K} S_{v, k, t}$ were also computed and compared to $S_t$ in order to verify that the choice of starting vertex $v$ had no impact on the outcome.
 
The whole process was repeated using a decreasing sequence of values of $p_d$, where each value was obtained by multiplying the previous value by a factor of $0.5$. Figure \ref{karate-simulation-figure} displays curves showing the average number of infected  vertices per round for $4$ different values of $p_d$. Each simulation was run with $p_b = 0.05$.

As predicted by Theorem \ref{NLDS-theorem}, when $p_d \ge p_b\lambda_1(G)$, the infection dies out. This is indicated by the lowest curve which tends to zero as the number of days increases. When $p_d < p_b\lambda_1(G)$ the infection lingers in the population as infected vertices recover and become infected again. This manifests itself in curves that sharply increase and then plateau at a constant positive number of infected vertices.

\begin{figure}[h]
\centering
\includegraphics[width=4in]{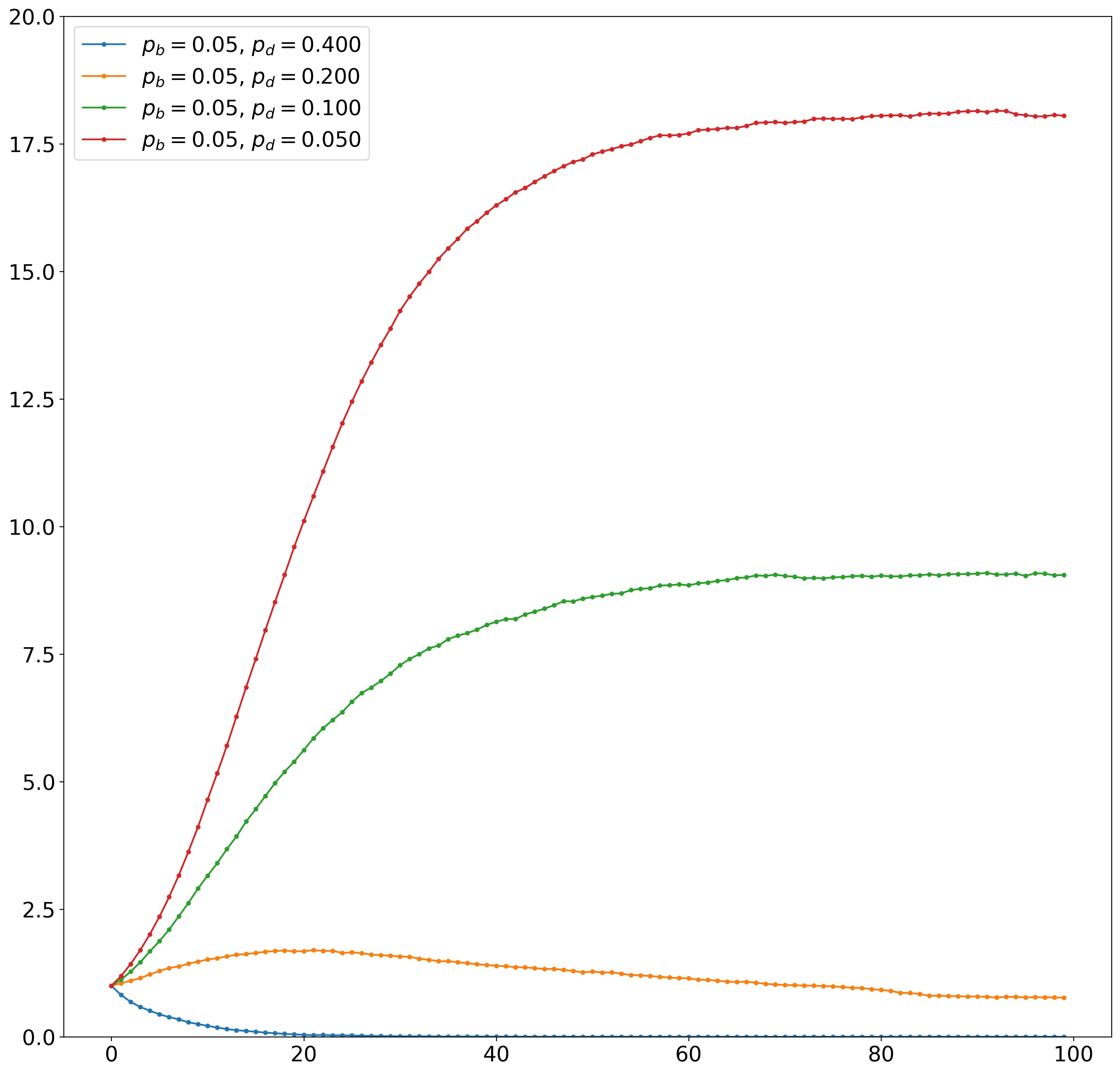}
\caption{Simulation results for karate graph with $p_b = 0.05$ and $p_d = 0.4, 0.2, 0.1$ and $0.05$. \label{karate-simulation-figure}}
\end{figure}
 
\subsection{Bounds on the largest eigenvalue} 
Let $G$ be a connected graph. Let $\Delta(G)$ denote the largest degree of $G$. Let $\chi(G)$ denote the chromatic number, that is the smallest number of colors required to color each vertex so that adjacent vertices receive different colors. Let $\bar{d}(G)$ denote the average degree if $G$. There are many results on upper and lower bounds of the largest eigenvalue involving well known graph parameters\cite{Cvet1990}. In \cite{Collatz1957} the observation was made that for a graph $G$ with $n\ge 2$ vertices and $m\ge 1$ edges $$2 \cos\frac{\pi}{n+1} \le \lambda_1(G)\le n-1,$$
and that equality for the lower bound holds if and only if $G$ is the path graph on $n$ vertices $P_n$ and for the upper bound if and only if $G$ is the complete graph on $n$ vertices $K_{n}$. It is also straightforward to see  that $$\bar{d}(G) \le \lambda_1(G)\le \Delta(G).$$ 
This was the starting point of the search for upper and lower bounds. In \cite{Nosa1970} it is shown that $\lambda_1(G) \ge \sqrt{\Delta(G)}$, where equality holds if and only if $G$ is the star graph $S_n$. Note that $K_{r,s}$ is the complete bipartite graph with $r$ vertices in one class and $s$ in the other, and $S_n = K_{1,n-1}$. In \cite{Wilf1967} it is shown  that $\lambda_1(G) \ge \chi (G) -1$, where equality holds if and only if $G$ is $K_n$ or the cycle on $n$ vertices $C_n$ with $n$ odd. These are just a few of the enormous variety of similar results.

\begin{theorem} \label {lowerbounds}  Let $G$ be a graph with $n\ge 2$ vertices and $m\ge 1$ edges. Then 
\begin{itemize}
\item [(i)] $\lambda_1(G) \ge \frac{2m}{n}$. Equality holds if and only if $G$ is regular of degree $\frac{2m}{n}$. 
\item [(ii)]  $\lambda_1(G) \ge \sqrt {\Delta(G)}$. Equality holds if and only if $G \cong S_n$.  
\item [(iii)]  $\lambda_1(G) \ge \chi (G) -1$. Equality holds if and only if $G \cong K_n$ or the odd cycle.
\end{itemize}
\end{theorem}
 
In the context of a biological contagion, there are three parameters that can be changed: the birth rate $p_b$, the death rate $p_d$, and  $\lambda_1(G)$. For example,  $p_b$ can be lowered by masking and washing hands frequently, $p_d$ can be increased by administering treatments to infected people to help them fight off the infection faster, and $\lambda_1(G)$ can be lowered by closing schools and asking people to work from home (reducing edges) or by vaccinating people (reducing the number of nodes).

Suppose we want to lower $p_b$, increase $p_d$, and lower $\lambda_1(G)$ to values that will guarantee that the virus will die out. How much is enough? In  \cite{Jam2006}, the authors argued that if we are unable to accurately compute $\lambda_1(G)$, then we are limited to computing upper bounds using parameters that we can approximate. Therefore, a sharper upper bound will yield a more accurate estimate of how much $p_b$ must be decreased or $p_d$ must be increased.  To this we add that a sharp lower bound on $\lambda_1(G)$ can also help guide decision making, as it indicates that insufficient changes to $p_b$ and $p_d$ are guaranteed by Theorem \ref{NLDS-theorem} to allow the infection to linger in the population.
 
A {\it clique} in a graph is a maximal induced   subgraph that is isomorphic to a complete graph.  The {\it clique number} of a graph $G$, $\omega(G)$, is the size of the largest  maximal clique. The following result indicates conditions under which an infection will be guaranteed to linger in the population, in terms of the average degree, largest degree and largest clique size in the graph, based on the predictions of Theorem \ref{NLDS-theorem}.
 
\begin{theorem} \label {lowerbounds-2} In an NLDS, where $p_b$ and $p_d$ are the birth rate and death rate of the virus,   respectively, the epidemic will linger in the population irrespective of the size of the initial outbreak of infection under the following conditions: 
\begin{itemize}
\item [(i)] $np_d < 2mp_b$.
\item [(ii)] $p_d <  p_b\sqrt{\Delta(G) }$. 
\item [(iii)] $p_d < p_b(\omega(G) - 1)$. 
\end{itemize}
\end{theorem}

\begin{proof} Theorem \ref{NLDS-theorem} implies that the epidemic will linger if $$p_d < \lambda_1(G)p_b.$$ If $np_d < 2mp_b$ then by Theorem \ref{lowerbounds} (i) $$p_d < \frac{2m}{n}p_b \le \lambda_1(G)p_b,$$ and therefore the epidemic will not die out.   Similarly, (ii) follows from Theorem \ref{lowerbounds} (ii). Since $\omega(G) < \chi(G)$, (iii) follows from Theorem \ref{lowerbounds} (iii).
\end{proof}

This result confirms the intuition that a social network with a large number of links between nodes, high degree nodes and large cliques increases $\lambda_1(G)$ and makes it harder for epidemics to die out.


\section {\bf  Spread Centrality}

A  vertex centrality measure  is an assignment of numbers to the vertices $\{v_1, v_2, \dots , v_n\}$ of a graph $G$ with the goal of ranking them in a manner relevant to some real world situation. The NLDS model of how a virus spreads suggests  a vertex centrality measure. Vaccinating a vertex may be viewed as deleting the vertex since the virus can no longer spread along its links to neighboring vertices. For each vertex $v_i$, compute the eigenvalues of the adjacency matrix $A(G-v_i)$ of the subgraph $G-v_i$, and write them in non-increasing order as $$\lambda_1 (G-v_i)\ge \lambda_2(G-v_i) \ge  \cdots \ge \lambda_{n-1}(G-v_i) \ge \lambda_n(G-v_i).$$ 

Focusing on the largest eigenvalue of $G-v_i$ gives a ranking of the vertices with the smaller numbers in the set
$\{\lambda_1(G-v_1),  \lambda_1(G-v_2), \dots , \lambda_1(G-v_n)\}$ being more desirable as candidates for vaccination. To be consistent with other centrality measures, we will consider the  difference between  $\lambda_1 (G)$ and $\lambda_1 (G-v_i)$ and adopt the usual largest to smallest ranking of vertices, with vertices ranking higher being more desirable. 
We use Cauchy's Interlacing Theorem \cite{Hwang2004} to ensure that the difference is always non-negative. 

\begin{theorem} \label{InterlacingTheorem} Let $A$ be an $n\times n$ real symmetric matrix and let $B$ be a first minor obtained from $A$ by deleting the $i$th row and $i$th column. Let $\alpha_1 \ge \alpha_2 \ge \dots \ge \alpha_n$ be the eigenvalues of $A$ and let $\beta_1 \ge \beta_2 \ge \dots \ge \beta_{n-1}$ be the eigenvalues of $B$. Then $$\alpha_1 \ge \beta_1 \ge \alpha_2 \ge \beta_2 \ge \dots \ge \alpha_{n-1} \ge \beta_{n-1} \ge \alpha_n.$$ \end{theorem}

Since $A(G)$ is a real symmetric matrix and $A(G-v)$ is obtained from $A(G)$ by deleting the $i$th row and column, Theorem \ref{InterlacingTheorem} implies that  $$\lambda_1 (G) \ge \lambda_1(G-v) \ge \lambda_2(G) \ge \lambda_2 (G-v) \ge \dots \ge \lambda_{n-1}(G) \ge \lambda_{n-1}(G-v) \ge \lambda_n(G).$$
Consequently, $\lambda_1 (G)-\lambda_1 (G-v)\ge 0$. We define the {\it spread centrality} of a vertex $v$  as  $$\sigma(v)=\lambda_1 (G)-\lambda_1 (G-v).$$

For example, consider the graph $G$ shown in Figure \ref{deckofvertexdeletions}. The largest eigenvalue of $G$, $\lambda_1(G)=2.48$ and  the vertex orbits are $\{v_1, v_2\}$, $\{v_3, v_5\}$, and $\{v_4\}$.

\begin{figure}[h]
$\vcenter{\hbox{\includegraphics[width=1in]{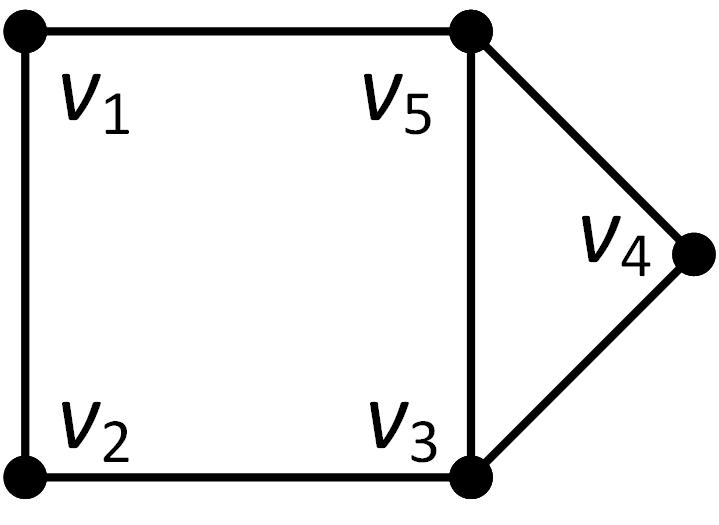}}}$\ \ 
$\tiny A=\bbordermatrix{
& v_1 & v_2 & v_3 & v_4 & v_5\cr
v_1 & 0 & 1 & 0 & 0 & 1 \cr
v_2 & 1 & 0 & 1 & 0 & 0 \cr
v_3 & 0 & 1 & 0 & 1 & 1 \cr
v_4 & 0 & 0 & 1 & 0 & 1 \cr
v_5 & 1 & 0 & 1 & 1 & 0
}
$
\end{figure}

Figure \ref{deckofvertexdeletions} shows the deck of vertex deletions of $G$ along with the largest eigenvalue of each graph in the deck. Table \ref{measures-table} shows the spread centrality of each vertex in contrast to the usual centrality measures.

\begin{figure}[h]
\centering
\includegraphics[width=5in]{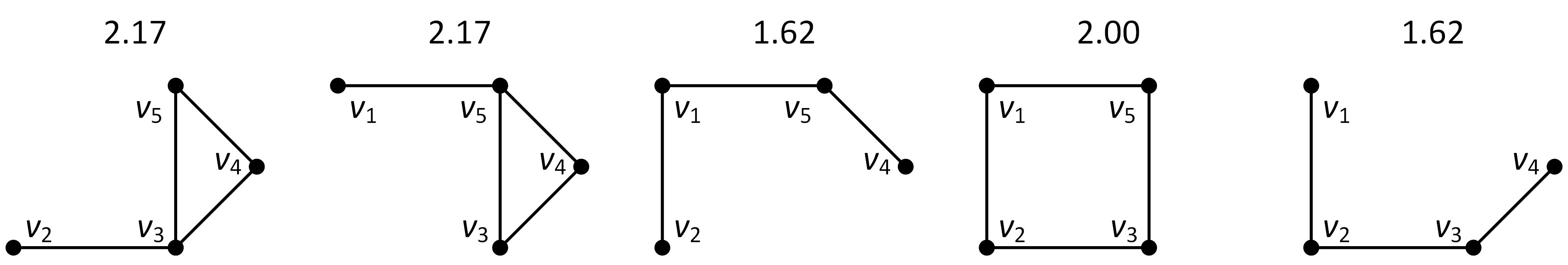}

\caption {Largest eigenvalues of the deck of vertex deletions of a graph.} \label{deckofvertexdeletions}
\end{figure}

\begin{table}[h]
\begin{tabular}{c|c|c|c|c|c}
Vertex & Spread & Degree & Closeness & Betweenness& Eigenvector  \\ \hline
$v_1$ & 2.48 - 2.17 = 0.31 & 2 & 0.17 & 0.5 & 0.36 \\
$v_2$ & 2.48 - 2.17 = 0.31 & 2 & 0.17 & 0.5 & 0.36 \\
$v_3$ & 2.48 - 1.62 = 0.86 & 3 & 0.2 & 1.5 & 0.53\\
$v_4$ & 2.48 - 2.00 = 0.48 & 2 & 0.17 & 0 & 0.43\\
$v_5$ & 2.48 - 1.62 = 0.86 & 3 & 0.2 & 1.5 &0.53  
\end{tabular} 
\caption{Spread and other centrality measures \label{measures-table}}
\end{table}

 In order to visualize the spread centralities of the vertices of a large graph, we can plot a heat map of the graph by assigning colors to vertices according to their spread centralities with red representing  highest spread centrality.
For example,  Figure \ref{karategraph} displays the spread centrality of the karate graph, which has largest eigenvalue $\lambda_1 (G) = 6.73$, and consequently  $p_d > 6.73p_b$.  In this example, $v_{16}$ has the smallest spread centrality and $v_{33}$ has the largest spread centrality, with values $0.01$ and $0.64$ respectively.  

\begin{figure}[h]
\centering
\includegraphics[width=3in]{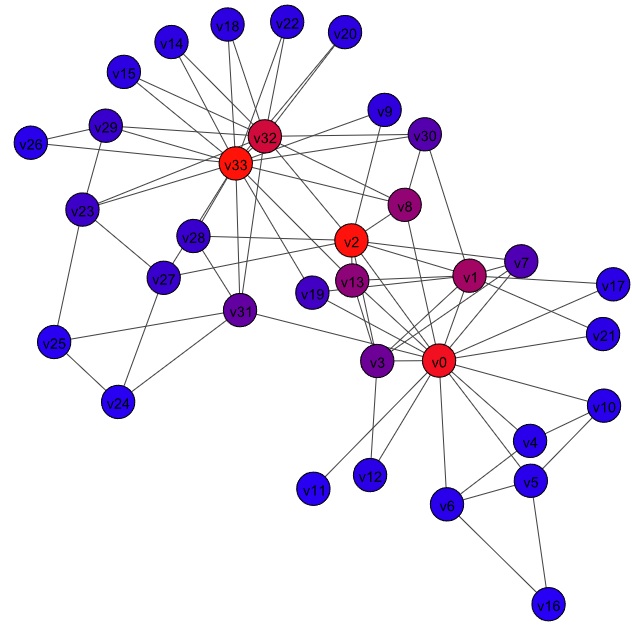}
\caption {Spread centrality for the karate graph \label{karategraph}}
\end{figure}
 
\subsection{Vaccination Strategies}
Let $G$ be a graph representing a social network in which an infection has to be controlled. Let $f$ be a function that accepts a graph with $n$ vertices as input and outputs a vector of length $n$ measuring the centralities of the vertices of the graph. For example, $f$ could output spread centralities or eigenvector centralities, as the case may be.

Suppose we have a fixed number $k$ of vaccine doses.  First observe that  
$$0 \le \sigma(v) \le \lambda_1(G).$$
The smaller the value of $\lambda_1(G-v)$ as compared to $\lambda_1(G)$, the larger the difference $\lambda_1(G)-\lambda_1(G-v)$,  which is desirable in order for the epidemic threshold $\tau(G-v)$ to be much larger than $\tau(G)$. It makes sense to vaccinate such  vertices first. We evaluated two ways to vaccinate (or remove) $k$  vertices: 

\noindent {\bf Method 1:} Select the $k$ vertices with the highest centrality according to $f$ and remove them. If there are ties, choose randomly among vertices at the bottom of the list. For example, suppose the top measures are $10$, $10$, $9$, $7$, $7$, and $7$ and $k = 5$. Then two of the vertices with score $7$ should be chosen randomly.
\bigskip

\noindent {\bf Method 2:} Remove the vertex with the highest centrality according to $f$, then recalculate $f(G-v)$ and remove the vertex with highest centrality, and so on sequentially until $k$ vertices are removed. If there are ties at any step, choose a vertex randomly from the vertices that have the highest centrality measure.

These methods were evaluated using two graphs, the previously mentioned karate graph with 34 vertices and 78 edges and the football network,\footnote{\url{http://www-personal.umich.edu/~mejn/netdata/}} described in \cite{GirvanNewman2002}, with 115 vertices and 613 edges. Eight vertices were removed from the karate graph and twenty-eight vertices from the football graph, or approximately one quarter of the vertices in each case.  The results, summarized in Table \ref{karate-football-methods-table},  clearly show that Method 2 is better. We examined several other large graphs to confirm this.
 
\begin{table}[h]
\begin{tabular}{|l|r|r|}
\hline
Graph & Method 1 & Method 2 \\
\hline
Karate & 2.62 & 2.17 \\
\hline
Football & 9.78 & 7.43 \\
\hline
\end{tabular}
\caption{Comparison of vaccination methods.}\label{karate-football-methods-table}
\end{table}

\subsection{Results on spread centrality}

A graph is {\it regular} if every vertex has the same degree.  A {\it tree} is a connected graph with no cycles and a {\it forest} is a graph with no cycles. A {\it leaf} is a degree 1 vertex.  

\begin{lemma} \label{lemma} If $G$ is a regular graph of degree $d$, then $\lambda_1(G)=d$, and the corresponding eigenvector is the vector of all ones.
\end{lemma}

\begin{lemma} \label{lemma2} If $G$ is the complete bipartite graph $K_{r,s}$, then $\lambda_1(G) = \sqrt{rs}$.
\end{lemma}
 
\begin{theorem} \label{examples-1} Let $K_n$, $C_n$, and $S_n$ be the complete graph, the cycle graph, and the star graph on $n$ vertices, respectively, and let $K_{t, t}$ be the complete bipartite graph with $t$ vertices in each vertex class.
\begin{enumerate} 
\item [(i)] For every vertex $v$ in $K_n$, $\sigma(v)=1$. 
\item [(ii)] For every vertex $v$ in $C_n$, $\sigma(v)=2(1-\cos\frac{\pi}{n})$.
\item [(iii)] For every vertex $v$ in $K_{t, t}$, $\sigma(v)=t-\sqrt{t(t-1)}$.
\item [(iv)] Let $v$ be the center vertex and $w$ be a leaf in $S_n$. Then $\sigma(v)=\sqrt{n-1}$ and $\sigma(w)=\sqrt{n-1}-\sqrt{n-2}$. 
\end{enumerate}
\end{theorem}
 
 \begin{proof} Since $K_n$, $C_n$, and $K_{t, t}$ are regular graphs of degree $n-1$, 2, and $t$, respectively, Lemma \ref{lemma} implies that $\lambda_1(K_n)=n-1$, $\lambda_1(C_n)=2$, and $\lambda_1(K_{t, t})=t$.
 
 For every $v$, $K_n- v \cong K_{n-1}$ and $\lambda_1(K_{n-1})=n-2$. Therefore  
 $\sigma(v)=1$. 
 
For every $v$, $C_n-v\cong P_{n-1}$ and  $\lambda_1(P_{n-1})= 2 \cos\frac{\pi}{n}$. Therefore  $\sigma(v)=2(1-\cos\frac{\pi}{n})$

For every $v$, $K_{t, t}-v\cong K_{t-1, t}$ and $\lambda_1(K_{t-1, t})=\sqrt{t(t-1)}$, by Lemma \ref{lemma2}. Therefore $\sigma(v)= t-\sqrt{t(t-1)}$.

Next, since $S_n = K_{1, n-1}$, Lemma \ref{lemma2} implies that $\lambda_1(S_n)=\sqrt{n-1}$. If $v$ is the center vertex, then $S_n-v$ is the isolated graph with $n-1$ vertices whose largest eigenvalue is 0. Therefore $\sigma(v)=\sqrt{n-1}$. If $w$ is a leaf, then $S_n-w \cong S_{n-1}$. So $\sigma(w)=\sqrt{n-1}-\sqrt{n-2}$. \end{proof}


\subsection{Comparision with other centrality measures} Using Spearman's rank correlation coefficient \cite{ZwilligerKikoska2000}, the spread centrality values for the vertices in the karate graph were compared with their degree, closeness, betweenness and eigenvector centralities. The results are shown in Figure \ref{correlations}.

\begin{figure}[h]
\centering
\includegraphics[width=6.5in]{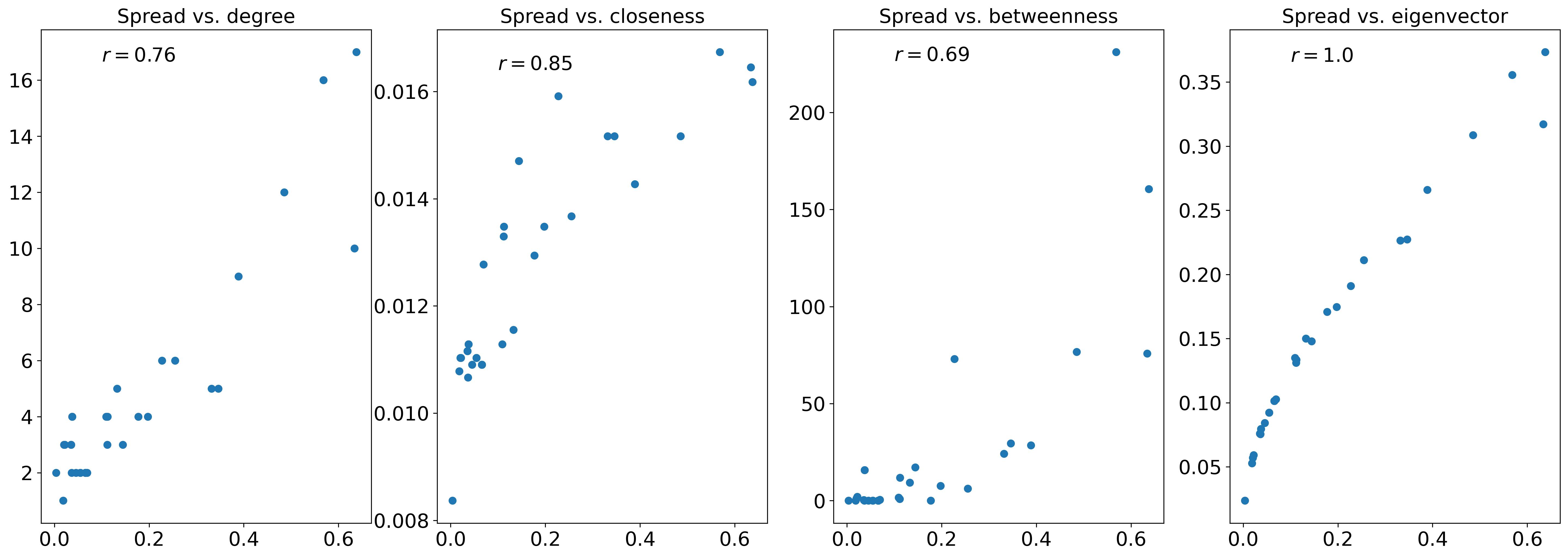}
\caption {Rank correlation between spread centrality and degree, closeness, betweenness, and eigenvector centralities  \label{correlations}}
\end{figure}
   
A high correlation with eigenvector centrality was noted. However, this is not always the case. For example, consider the  regular graph $G$ shown in Figure \ref{regular-graph-example}. Theorem \ref{examples-1}(vi) implies that $\lambda_1(G)=3$ and its eigenvector is the vector of all ones. Thus each vector has the same eigenvector centrality. However, the spread centrality is not the same. Vertices $v_1$, $v_3$, $v_5$, and $v_7$ have spread centrality  0.29 and vertices $v_2$, $v_4$, $v_6$, and $v_8$ have spread centrality 0.24.

\begin{figure}[h]
\centering
\includegraphics[width=1.5in]{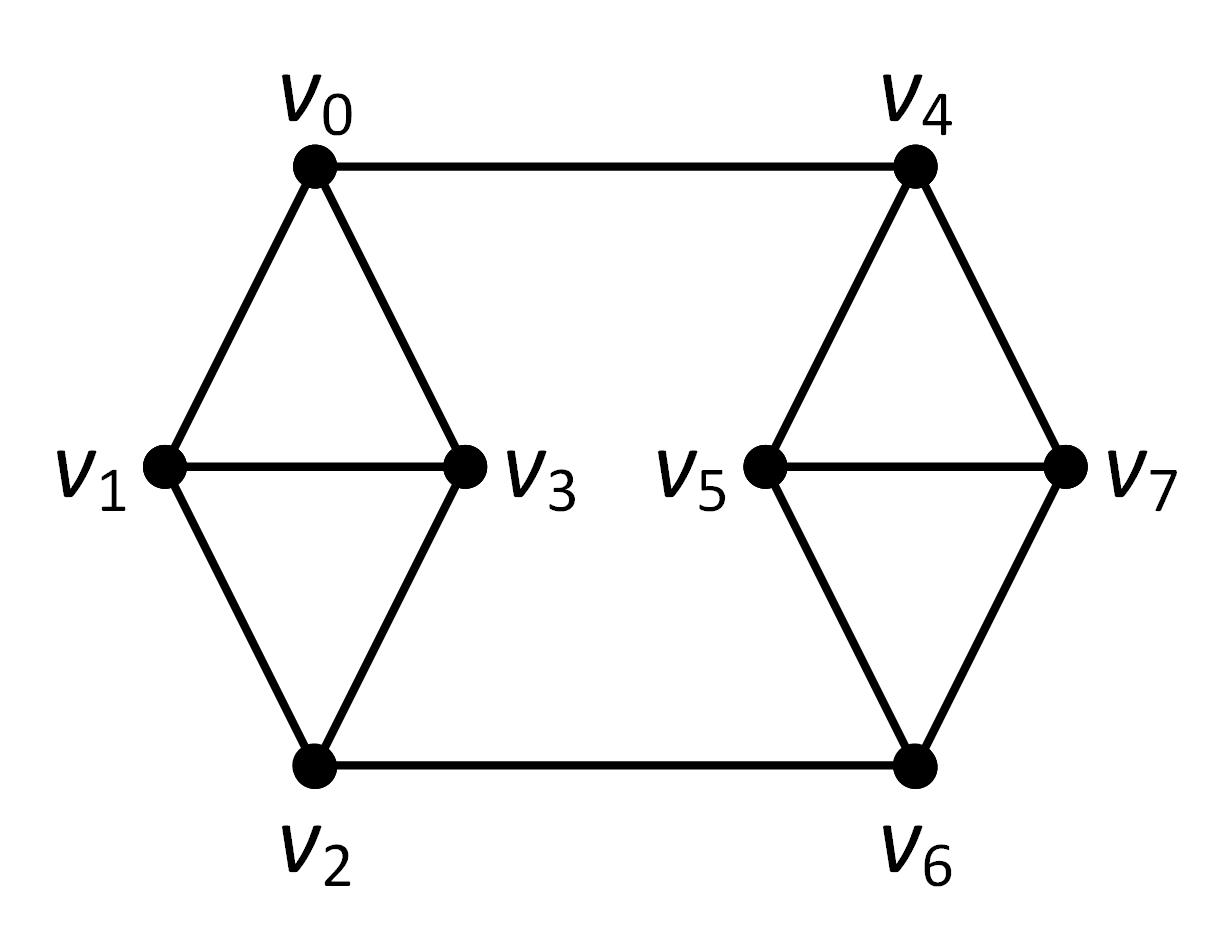}
\caption { A regular graph \label{regular-graph-example}}
\end{figure}

Eigenvector centrality ranks vertices according to their corresponding values in the eigenvector of the largest eigenvalue of the adjacency matrix. Canright and Engo-Monsen argued in \cite{Can2006} that eigenvector centrality is a good measure of how a disease propagates from node to node in a graph. Our experimental analysis confirms their premise. Theorem \ref{NLDS-theorem} states that the spread of a disease depends on the largest eigenvalue of the adjacency matrix. Our definition of spread centrality, based on Theorem \ref{NLDS-theorem}, is defined as the difference between $\lambda_1(G)$ and $\lambda_1(G-v)$ and therefore directly aims to select vertices whose removal will impact spread most significantly. The fact that such as strong correlation exists between these two parameters will form the basis of future work.

{\bf Acknowledgements:} The authors thank Robert Kingan for help with the simulation.

\vfill
\eject
 
\section* {\bf Appendix A}

The simulation algorithm described below assumes the existence of a function \textproc{Rand} which returns a random number drawn uniformly from the interval $[0, 1]$. Let $s_{w, t}$ denote the state of  vertex $w$ at time $t$, which may be either $0$ (susceptible) or $1$ (infected).

\small
\begin{algorithm}
\caption{Disease propagation simulation algorithm}
\begin{algorithmic}[1]
\Procedure{SimulateDisease}{Graph $G$, $0 < p_b < 1$, $0 < p_d < 1$, $K > 0$, $T > 0$}
\For {$v \in V(G)$}
	\For{$k$ in $1, 2, \dots, K$}
		\For{$w \in V(G)$}
			\If{$w = v$}
				\State $s_{w, t} \gets 1$
			\Else
				\State $s_{w, t} \gets 0$
			\EndIf
		\EndFor
		\For{$t$ in $2, 3, \dots, T$}
			\For{$w \in V(G)$}
				\If{$s_{w, t-1} = 0$}
					\State{$i \gets \sum_{x \in N(w)}s_{x, t-1}$}
					\State{$s_{w, t} \gets 0$}
					\For{$j$ in $1, \dots, i$}
						\State{$r \gets \Call{Rand}{\ }$}
						\Comment Draw a random number in the range $[0, 1]$.
						\If{$r < p_b$}
							\Comment If the number is less than $p_b$, 
							\State{$s_{w, t} \gets 1$}
							\Comment vertex $w$ becomes infected.
							\Break
						\EndIf
					\EndFor
				\Else
					\Comment An infected  vertex recovers with probability $p_d$.
					\State{$r \gets \Call{Rand}{\ }$}
					\If{$r < p_b$}
						\State{$s_{w, t} \gets 0$}
					\Else
						\State{$s_{w, t} \gets 1$}
					\EndIf
				\EndIf
			\EndFor
		\EndFor
	\EndFor
\EndFor
\State{$n \gets |V(G)|$}
\For{$t$ in $1, 2, \dots, T$}
	\State{$S_t \gets \frac{1}{nK} \sum_{v \in V(G)} \sum_{1 \le k \le K} s_{v, k, t}$}
\EndFor

\State{OUTPUT $\{S_t | 1 \le t \le T\}$}

\EndProcedure
\end{algorithmic}
\end{algorithm}

\normalsize
\vfill 
\eject


\vfill \eject


\end {document}